\documentclass{iau}
\usepackage{graphicx}
\usepackage{amssymb}
\usepackage{rotating}
\usepackage{amsmath,wasysym}

\title[Anelastic scaling laws] 
{Bridging planets and stars using scaling laws in anelastic spherical shell dynamos}

\author[Yadav et al.]   
{R. K. Yadav$^{1,2}$, T. Gastine$^1$, U. R. Christensen$^1$, and L. Duarte$^{1,3}$}

\affiliation{$^1$Max-Planck-Institut f\"ur Sonnensystemforschung,  37191 Katlenburg-Lindau, Germany \\[\affilskip]
$^2$Institut f\"ur Astrophysik, Georg-August-Universit\"at,  37077 G\"ottingen, Germany \\[\affilskip]
$^3$Technische Universit\"at Braunschweig, Germany}

\pubyear{2013}
\volume{302}  
\pagerange{119--126}
\jname{Magnetic fields throughout stellar evolution}
\editors{M. Jardine, P. Petit \& H. Spruit, eds.}
\begin{document}

\maketitle

\begin{abstract}
Dynamos operating in the interiors of rapidly rotating planets and low-mass stars might belong to a similar category where rotation plays a vital role. We quantify this similarity using scaling laws. We analyse direct numerical simulations of Boussinesq and anelastic spherical shell dynamos. These dynamos represent simplified models which span from Earth-like planets to rapidly rotating low-mass stars. We find that magnetic field and velocity in these dynamos are related to the available buoyancy power via a simple power law which holds over wide variety of control parameters.
\end{abstract}

{\bf \em Introduction}: In the last decade or so some qualitative agreement has been found in geodynamo simulations and observations (see e.g. \cite[Jones 2011]{Jones2011a}). However, a direct and quantitative comparison of simulations and observations is not possible because of the large diffusivities used in numerical simulations as compared to the astrophysical values. To better connect numerical simulations with observations it is thus of great importance to find out generic scaling laws which are valid for both.

\cite[Christensen \& Aubert (2006)]{Christensen2006} found consistent scaling laws for magnetic field and velocity as a function of the available buoyancy power in Boussinesq spherical-shell dynamo simulations. \cite[Christensen et al. (2009)]{Christensen2009} extended the magnetic field scaling law to physically relevant parameter regime and found good agreement with magnetic field observed on Earth, Jupiter, and some rapidly rotating low-mass stars.

Numerical scaling studies mentioned above ignored compressibility as they were geared to model dynamos operating in liquid metal interiors of Earth like planets. Giant planets and low-mass stars on the other hand might have highly compressible interiors with radially varying diffusivities (\cite[French et al. 2012]{French2012}). Assessing the effect of compressibility on various scaling laws in dynamos is very important to better understand the dynamo mechanism in giant planets and rapidly rotating low-mass stars.

{\bf \em Results}: In recent years few extensive parameter-studies have been performed to study various aspects of compressible dynamos using the anelastic approximation (\cite[Gastine et al. 2012]{Gastine2012dyn}; \cite[Gastine et al. 2013]{Gastine2013}; \cite[Duarte et al. 2013]{Duarte2012}). We use this dataset along with Boussinesq dynamos (\cite[Yadav et al. 2013a]{Yadav2013a}) to explore the scaling of different quantities. Lorentz number $Lo=(\int{(\mathbf{B}\cdot\mathbf{B})}\,dV/\int{\tilde{\rho}\,dV})^{1/2}$ represents the mean magnetic field and convective Rossby number $Ro_{conv}=(\frac{1}{V}\int{(\mathbf{u}_{na}\cdot\mathbf{u}_{na})}\,dV)^{1/2}$ represents the mean convective velocity, where $\mathbf{B}$, $\mathbf{u}_{na}$ is non-dimensional magnetic field and non-axisymmetric velocity, respectively, and $\tilde{\rho}$ is radially varying density. Note that such averaging can be described as magnetic and kinetic energy per-unit-mass, which is more appropriate for density varying interiors. 

Despite radially-varying properties, $Lo$ and $Ro_{conv}$ still scale consistently as a function of the available buoyancy power per-unit-mass $P$ as shown in Fig.~1. Empirical power-law describing magnetic field scaling is $Lo/\sqrt{f_{ohm}}=c\,P^{0.33}\,{P_m}^{0.1}$, with $c=0.9$ for dipolar dynamos and $c=0.7$ for multipolar dynamos, and convective velocity is $Ro_{conv}=1.6\,P^{0.42}\,P_m^{-0.08}$ (\cite[Yadav et al. 2013b]{Yadav2013b}). The $Lo$ scaling requires $f_{ohm}$ (fraction of total energy lost as ohmic heating) for a consistent scaling behaviour. Both scalings also require inclusion of $P_{m}$ (magnetic Prandtl number) for optimum fit quality. However, such $P_{m}$ dependence might be only due to the rather large diffusivities employed in present numerical models and may not be important for natural objects where $P_{m}\ll 1$ (\cite[Christensen et al. 2010]{Christensen2010}).

In summary, we generalize the scaling laws found in earlier studies to compressible dynamos and support the hypothesis that magnetic field and velocity are related to the available buoyancy power by power-laws in dynamos. Decent observational evidence exists for the magnetic field scaling (\cite[Christensen et al. 2009]{Christensen2009}) but comparison of velocity scaling with observations has not been possible so far (except for Earth's core).

\begin{figure}
\begin{center}
\includegraphics[scale=0.32]{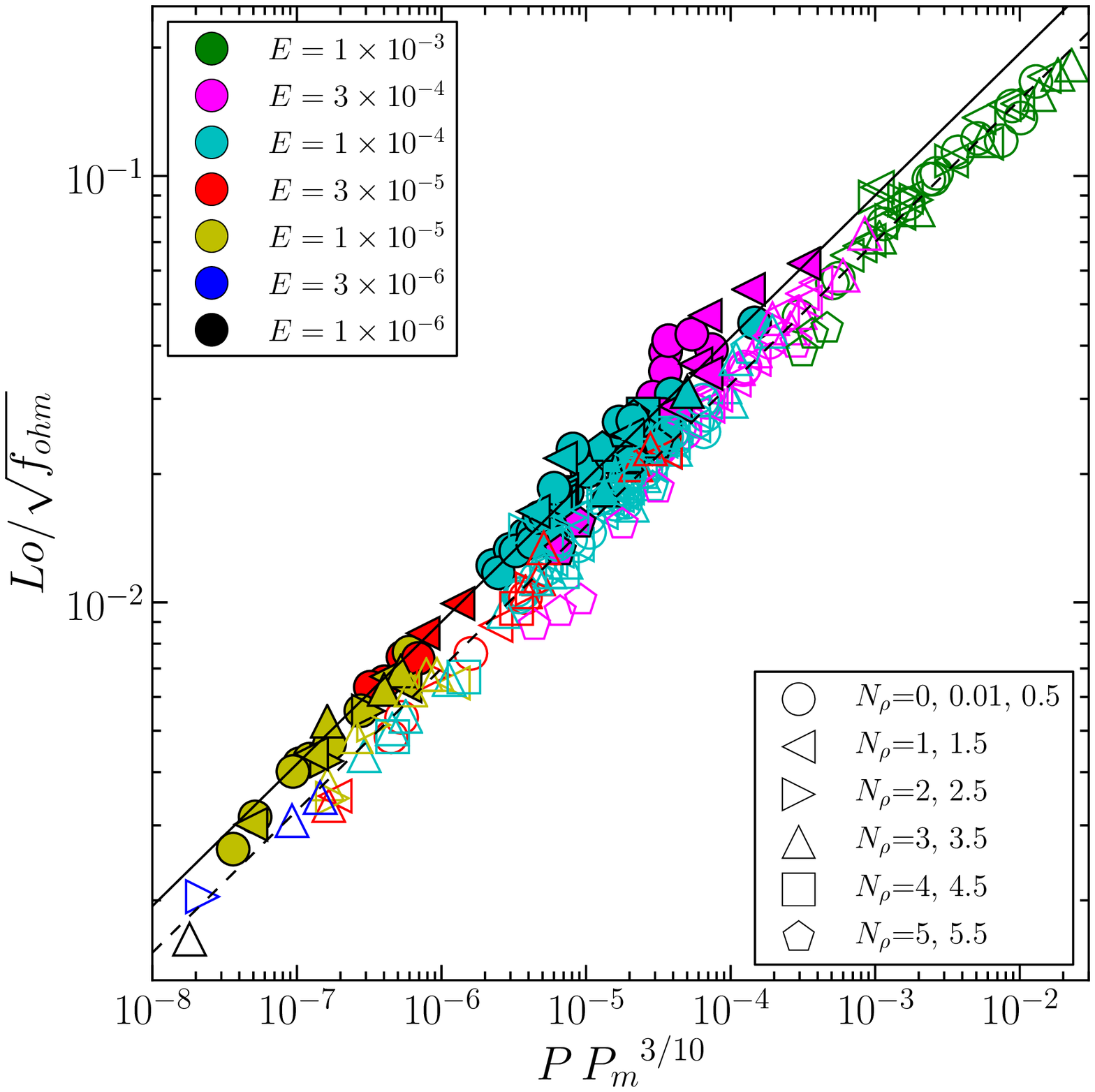} \hspace{5 mm} \includegraphics[scale=0.32]{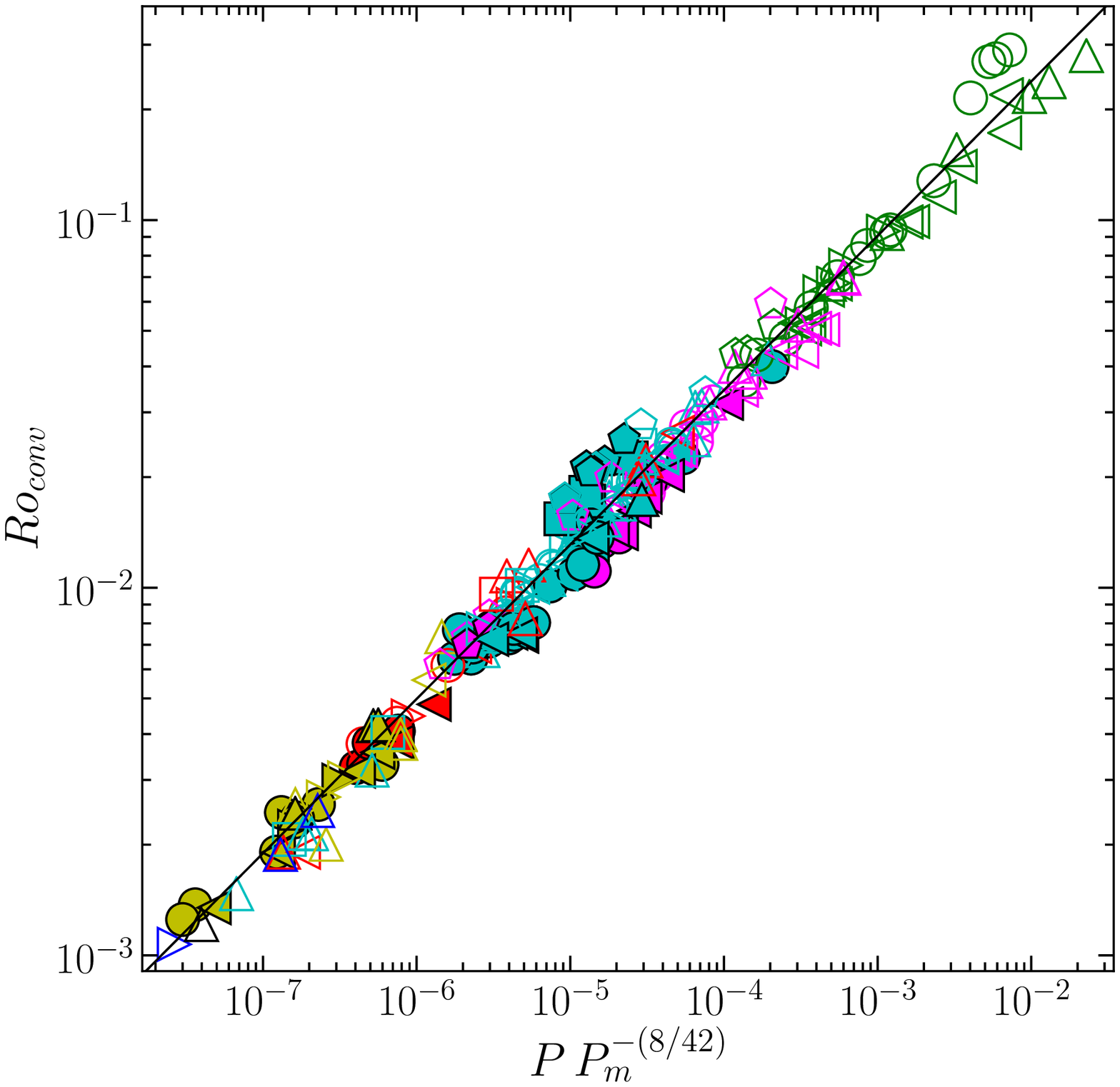} 
\end{center}
\caption{Scaling of non-dimensional magnetic field $Lo$ (left panel) and convective velocity $Ro_{conv}$ (right) as a function of the buoyancy power per-unit-mass $P$. Filled (empty) symbols are dipolar (multipolar) dynamos. Symbol color represents Ekman number $E$ (right panel; top left legend) and symbol shape represents number of density scale heights $N_{\rho}$ in the convecting fluid shell (right panel; bottom right legend).}
\end{figure}

{\bf \em Acknowledgements}: We acknowledge funding from the DFG through Project SFB 963/A17 and through the special priority program 1488 (PlanetMag). Simulations were run on GWDG and HLRN computing facilities.

\end{document}